\documentclass[aps,pre,twocolumn,groupedaddress,floatfix]{revtex4-1}
\usepackage{color}
\usepackage{tabularx}
\usepackage{epsfig}
\usepackage{amsmath}
\usepackage{graphicx}

\begin{document}

\title{The Ising universality class in dimension three : corrections to scaling}

\author{P. H.~Lundow} \affiliation {Department of Mathematics and
  Mathematical Statistics, Ume{\aa} University, SE-901 87 Ume{\aa},
  Sweden}

\author{I. A.~Campbell} \affiliation{Laboratoire Charles Coulomb
  (L2C), UMR 5221 CNRS-Universit\'e de Montpellier, Montpellier,
  F-France.}

\date{\today}

\begin{abstract}
Simulation data are analyzed for four 3D spin-$1/2$ Ising models: on
the FCC lattice, the BCC lattice, the SC lattice and the Diamond
lattice. The observables studied are the susceptibility, the reduced
second moment correlation length, and the normalized Binder
cumulant. From measurements covering the entire paramagnetic
temperature regime the corrections to scaling are estimated.  We
conclude that a correction term having an exponent which is consistent
within the statistics with the bootstrap value of the universal
subleading thermal confluent correction exponent, $\theta_{2} \sim
2.454(3)$, is almost always present with a significant amplitude. In
all four models, for the normalized Binder cumulant the leading
confluent correction term has zero amplitude. This implies that the
universal ratio of leading confluent correction amplitudes
$a_{\chi_{4}}/a_{\chi} = 2$ in the 3D Ising universality class.
\end{abstract}


\maketitle

\section{Introduction}\label{sec:I}
The development of the "conformal bootstrap" approach has led to a
major step forward in understanding the canonical Ising universality
class in dimension three \cite{showk:12, showk:14, kos:14, simmons:15,
  gliozzi:15, nakayama:16}. The principle universal critical
exponents: the correlation length exponent $\nu = 0.62999(5)$, the
anomalous dimension $\eta = 0.03631(3)$ (and so the susceptibility
exponent $\gamma = (2-\eta)\nu = 1.23710(12)$ and the specific heat
exponent $\alpha= D\nu-2 = 0.11003(15)$ ) and the leading confluent
correction exponent $\omega =0.8303(18)$ (and so the thermal confluent
correction $\theta= \omega\nu = 0.5231(11)$) are established to high
precision. Earlier high temperature scaling expansion (HTSE)
\cite{butera:02,campostrini:02} and numerical simulation
\cite{deng:03,hasenbusch:10} values are slightly less accurate than
but fully consistent with the bootstrap results.

The subleading conformal correction exponent, which is also universal,
was however never directly estimated in earlier work but was simply
assumed to be $\omega_{2} = 1.67(11)$ (and so the thermal confluent
correction exponent $\theta_{2}= \omega_{2}\nu = 1.05(7)$) following
Ref.~\cite{newman:84}. For this exponent the bootstrap estimates are
dramatically higher : $\omega_{2} \sim 4.3$, and so $\theta_{2} \sim
2.7$ \cite{showk:12,showk:14,gliozzi:15,nakayama:16}; the most recent
bootstrap calculation, Ref.~\cite{simmons:17} Table 2, provides a
high precision estimate for the $\epsilon^{''}$ stable operator
parameter $\Delta = 6.8956(43)$ which can be translated to give the
second thermal conformal correction exponent $\theta_{2} =
2.454(3)$. Here we examine extensive numerical and HTSE data on the 3D
Ising universality class in the light of this result.

Historically, the standard thermal scaling variable in Ising models
was $\tau = (1-\beta/\beta_{c})$ where $\beta$ is the inverse
temperature $1/T$. $\tau$ was used initially by Domb and Sykes in 1962
\cite{domb:62}, by Wegner for his RGT expansion in 1972
\cite{wegner:72}, and has been in continuous use in HTSE analyses ever
since (see for instance Ref.~\cite{butera:02}). $\tau$ varies from $0$
at criticality to $1$ at infinite temperature, with no divergence, so
by using $\tau$ as the thermal scaling variable HTSE and simulation
data can be analysed in detail from criticality to infinite
temperature with empirical scaling analyses in terms of temperature
dependent effective exponents such as $\gamma(\tau)$, as has already
been demonstrated in various specific Ising and ISG models,
e.g. \cite{butera:02,campbell:06,campbell:08,lundow:11a,lundow:11,berche:08}.

The Wegner expansion for the susceptibility can be written (see for
instance Ref.~\cite{privman:91})
\begin{equation}
  \chi(\tau) = C_{\chi}\tau^{-\gamma}\left(1+a\tau^{\theta}+ b\tau +
  c\tau^{\theta_{2}} +\cdots\right)
  \label{chiweg}
\end{equation}
where only the three leading correction terms are written explicitly;
the exponents are universal but the critical amplitude $C_{\chi}$ and
the correction amplitudes $a$, $b$, $c$ are not, though conformal
correction amplitude ratios such as $a_{\chi}/a_{\xi}$ are universal
\cite{ferer:77,chang:80}. The first correction term is the leading
confluent correction, the second is the leading analytic correction,
and the third one is the subleading confluent correction. Equivalent
expressions can be written for other thermodynamic variables $Q(\tau)$
\cite{butera:02}. There is little in the way of {\it a priori}
guidelines as to expected critical amplitudes or correction amplitudes
for specific models.  A forbidding list of further potential
correction terms is indicated by Privman {\it et al.}
\cite{privman:91}, with "minor" terms in $\tau^2$, $\tau^{\theta+1}$,
$\tau^{2\theta}$, etc. As Eq.~\eqref{chiweg} can be written
\begin{eqnarray}
  \chi(\tau) &=& C_{\chi}\tau^{-\gamma}\left(1 + a\tau^{\theta}[1 +
    a_{1}\tau^{\theta} + a_{2}\tau + \cdots]\right. \nonumber\\ &+&
  b\tau[1 + b_{1}\tau^{\theta} + b_{2}\tau + \cdots]
  \nonumber\\ &+& c\tau^{\theta_{2}}[1 + c_{1}^{\theta} +
    c_{2}\tau + \cdots] \: + \left.\cdots\right)
    \label{chiwegB}
\end{eqnarray}
all the "minor" terms can be considered as corrections to
corrections. Leading correction amplitudes $a,b,c$ turn out to be
typically $0.10$ or less, so a plausible assumption is that the
"correction to correction" terms have very small amplitudes, $\sim
0.001$. Indeed Ref.~\cite{campostrini:02} states "several ["minor"]
corrections ... apparently conspire to give a uniformly small
correction", and Ref.~\cite{hasenbusch:10} states "We estimate the
error caused by ["minor"] correction terms that are not included by
comparing the results obtained by using different ans\"{a}tze and
... by fitting different quantities." In the present three-term
analyses including only the three leading correction terms in
Eq.~\eqref{chiweg} with exponents $\theta = 0.523$, $1$, and
$\theta_{2} = 2.45$, the "minor" correction having exponent
$\tau^{2\theta} \sim 1.04$ would be confounded with the leading
analytic correction term to give an effective amplitude. If any
neglected "minor" contributions from corrections having exponents of
the order of $\sim 1.5$ have significant amplitudes they would be
visible as perturbations to the three-term fits. No evidence for such
perturbations has been seen in the fits described below, so "minor"
correction terms will be considered negligible.

Although the Wegner expression was initially introduced for improving
asymptotic scaling analyses "near the critical point", the temperature
dependence of effective exponents such as $\gamma(\tau,L)=
-\partial\ln \chi(\tau,L)/\partial\ln\tau$ can be readily measured up
to infinite temperatures by HTSE or numerically, with HTSE data
becoming essentially exact for high $\tau$ if $\beta_{c}$ is well
known (see \cite{butera:02}).  A Wegner correction term with an
exponent considerably higher than $1$ will influence the data
significantly only at high $\tau$ so in practice the "near the
critical point" condition must be relaxed. Thus, in
Ref.~\cite{butera:02} Figs.~14 and 16 all the effective susceptibility
exponent $\gamma(\tau)$ curves for SC and BCC lattices and for spins
from $S=1/2$ to $S=\infty$ can be seen by inspection to have high
temperature upturns, consistently indicative of a negative high
exponent correction term of strong amplitude.

\section{Scaling analyses}\label{sec:II}
Traditional analyses of simulation data in general focus either on
finite size scaling (FSS) at the critical temperature, or on scaling
as a function of temperature using the thermal scaling parameter
$t=(T-T_c)/T_c$. This approach follows the choice made by K.~Wilson
who expressed Renormalization Group Theory (RGT) in terms of
$t$. However $t$ diverges at high temperatures, so $t$ scaling can
obviously only be used in the near-critical regime. In consquence,
according to the conventional wisdom critical exponents and intrinsic
correction terms can only be estimated from numerical measurements
using high precision simulations close to $T_c$ for large sample sizes
$L$. As high exponent correction terms only become important at
temperatures well above criticality, the $t$ scaling approach can be
ruled out for obtaining numerical or HTSE evidence concerning
subleading conformal corrections with exponent $\theta_{2} \sim 2.45$.

(It is underlined in Ref.~\cite{campostrini:02} that it can be
possible to modify a model Hamiltonian so as to give an "improved"
Hamiltonian, where the amplitude of the leading conformal correction
term in $\tau^{\theta}$ becomes zero, for all thermodynamic
variables. All the "minor" correction terms containing factors
$\tau^{\theta}$ will then also be suppressed simultaneously, again for
all observables. )

The Wegner expansion is for infinite samples, but holds also for
finite-size samples in the regime where $L \gg \xi(\tau)$; in practice
the rule $L > 7\xi(\tau)$ is sufficient (see for instance
Ref.~\cite{kim:96}) . When this condition holds, observable
$Q(\tau,L)$ data for all $L$ correspond to the infinite-$L$ limit
$Q(\tau,\infty)$ and so data for all $L$ coincide. Explicit
comparisons between $L$ and $\xi(\tau)$ are generally not needed to
establish where the ThL limit holds, as the ThL regime can be
recognized by inspection of data plots for $\chi(\tau,L)$ against
$\tau$, or equivalent plots for other observables $Q(\tau,L)$. Once
the ThL plots for $Q(\tau,\infty)$ and $\xi(\tau,\infty)$ including
the thermal scaling corrections have been established, $Q(\tau,L)$
data for all $L$ and all $\tau$ can be concatenated through the
Privman-Fisher finite-size scaling rule \cite{privman:84},
$Q(\tau,L)/Q(\tau,\infty) = F[L/\xi(\tau,\infty)]$.

It can be noted that at infinite temperature for spin $S=1/2$ models
the susceptibility $\chi(\tau=1,L) \equiv 1$. It was pointed out in
Ref.~\cite{campbell:06} that Wegner expressions for other observables
$Q(\tau)$ only take up strictly the susceptibility form with
non-diverging correction amplitudes if the observable is normalized
such that at infinite temperature $Q_{n}(\tau=1)=1 $. (Note that with
three correction terms this rule imposes the closure condition
$C(1+a+b+c)=1$). As well as the susceptibility we will study data on
the near-neighbor second-moment correlation length $\xi(\tau)$ and on
the Binder cumulant $g(\tau)$. $\xi(\tau)$ always tends to
$\xi(\tau=1)= 0$ at infinite temperature; from the general HTSE series
\cite{butera:02a} the leading $S=1/2$ series term for the
nearest-neighbor second-moment correlation is $\mu_2(\beta) =
z\,\beta$ where $z$ is the number of near neighbors. So when
$\xi(\tau)$ is defined appropriately for the lattice being considered,
the reduced correlation length $\xi(\tau)/\beta^{1/2}$ will have an
exact high temperature limit $\xi(\tau)/\beta^{1/2} = 1$, and an
unaltered critical exponent, as carefully explained in
Refs.~\cite{campbell:06,campbell:08}. This reduced correlation length
has a Wegner temperature dependence
\begin{equation}
  \frac{\xi(\tau)}{\beta^{1/2}} =
  C_{\xi}\tau^{-\nu}\left(1+a_{\xi}\tau^{\theta}+ b_{\xi}\tau +
  c_{\xi}\tau^{\theta_{2}} +\cdots\right)
  \label{xiscal}
\end{equation}
with the universal critical exponent $\nu$ and with correction
amplitudes which, as will be seen below, turn out to be weak so the
effective ThL reduced correlation-length exponent
$\nu(\tau)=\partial\ln[\xi(\tau,L)/\beta^{1/2}]/\partial\ln\tau$
varies little over the entire paramagnetic temperature range.

Assuming hyperscaling, the critical exponent for the second field
derivative of the susceptibility $\chi_{4}(\tau)$ (also called the
non-linear susceptibility) is Ref.~\cite{butera:02}
\begin{equation}
  \gamma_{4}=\gamma +2\Delta_{\mathrm{gap}}  = D\nu + 2\gamma
  \label{gam4}
\end{equation}
$\chi_{4}$ in a cubic lattice is directly related to the Binder
cumulant through
\begin{equation}
  2g(\tau,L) = \frac{-\chi_{4}}{L^D\chi^{2}} = \frac{3\langle
    m^2\rangle^2 - \langle m^4\rangle}{\langle m^2\rangle^2}
\end{equation}
see Eq.~{(10.2)} of Ref.~\cite{privman:91}.  Thus in the ThL regime
the normalized Binder cumulant $L^{D}g(\tau,L) \equiv
-\chi_{4}(\tau,L)/(2\chi(\tau,L)^2)$ scales with a critical exponent
$(D\nu + 2\gamma) - 2\gamma = D\nu$.  In any $S = 1/2$ Ising system
the infinite-temperature (i.e. independent spin) limit for the Binder
cumulant is $g(0,N) \equiv 1/N$, where $N$ is the number of spins. As
$N = L^{D}$ for a cubic lattice with $L$ defined appropriately, at
infinite temperature $L^{D}g(\tau,L)\equiv 1$. Thus the 3D normalized
Binder cumulant $L^{3}g(\tau)$ also obeys the high-temperature limit
rule for normalized observables introduced above, and the appropriate
Wegner expression is
\begin{equation}
  L^{3}g(\tau,L) = C_{g}\tau^{-3\nu}\left(1+a_{g}\tau^{\theta}+
  b_{g}\tau + c_{g}\tau^{\theta_{2}} +\cdots\right)
  \label{gscal}
\end{equation}

\section{Simulations and analyses}\label{sec:III}
We will present data measured over the entire range from criticality
to infinite temperature for spin $S=1/2$ Ising models on cubic, body
centered cubic, simple cubic, and diamond lattices presented in
decreasing order of the number of near neighbors. Most of the data
were originally generated for the critical regime analyses of
Ref.~\cite{haggkvist:07,lundow:09}, where the critical temperatures
and critical exponents were estimated. The susceptibility up to high
temperatures for these lattices (together with others) was presented
in Ref.~\cite{lundow:11a} where it was shown that the "crossover"
behavior to a high-temperature scaling regime claimed in
Refs.~\cite{luijten:97,luijten:99} was an artefact due to the use by
these authors of $t$ as the thermal scaling variable. A detailed
analysis of the simple cubic lattice susceptibility and specific heat
data along the lines of the present work was described in
Ref.~\cite{lundow:11}.

For the four cubic lattices we analyse the data for the susceptibility
$\chi(\tau,L)$, the reduced second-moment correlation length
$\xi(\tau,L)/\beta^{1/2}$, and the normalized Binder cumulant
$L^{3}g(\tau,L)$ over the whole paramagnetic temperature regime
assuming that in each case the temperature dependence of the
observable follows a Wegner expression with a limited set of
three (at most) correction terms, i.e.,
\begin{equation}
  Q(\tau)= C_{q}(\tau)\tau^{-\lambda_{q}}\left(1 + a_{q}\tau^{\theta}
  + b_{q}\tau + c_{q}\tau^{\mu_{q}}\right)
\end{equation}
which is the generalization of Eq.~\eqref{chiweg}, with $\lambda_{q}$
standing for the known bootstrap critical exponents : $\gamma$, $\nu$
and $3\nu$ respectively, and $\mu_{q}$ is the bootstrap subleading
conformal correction exponent from Ref.~\cite{simmons:17}. The
amplitudes are estimated from the fits. All "minor" correction terms
are assumed to have negligible amplitudes. In view of the number of
fit parameters we do not attempt to estimate the errors in the
individual correction amplitudes. Our final aim is to show that the
data are consistent with the presence in all the data sets of
correction terms having a unique value for $\mu$ (identified with
$\theta_{2} \sim 2.45$) and significant amplitudes. (Exceptionally
there might be evidence for a further high-order correction term which
could be ascribed to corrections with the further exponent
$\theta_{3}$).

To estimate critical amplitudes and the corrections to scaling,
simulation data and HTSE data when available can be displayed over the
entire paramagnetic temperature regime as $y(\tau) =
Q(\tau,L)\tau^{\lambda_{q}}$ against $x(\tau) = \tau^{\theta}$. When
the leading correction term is the confluent correction with exponent
$\tau^{\theta}$ this plot is linear at small $x(\tau)$ and the second,
analytic, term is nearly proportional to $x(\tau)^2$. This display is
appropriate for all the susceptibility and normalized correlation
length data.  However, we will see that in this Ising universality
class, for the normalized Binder cumulant the leading confluent
correction term is missing so the appropriate plot becomes $y(\tau)$
against $\tau$.  In addition for all observables the data can be
displayed in the form of effective temperature-dependent exponents
$\lambda_{q}(\tau) = -\partial\ln Q(\tau,L)/\partial\ln\tau$ (see
Ref.~\cite{butera:02}).

In the latter form of display with the three correction term
expression above one has
\begin{equation}
  \lambda_{q}(\tau) = \lambda_{q} - \frac{
    a_{q}\theta\tau^{\theta}+b_{q}\tau+c_{q}\mu\tau^{\mu}}{
    1 + a_{q}\tau^{\theta} +b_{q}\tau +c_{q}\tau^{\mu}}
\end{equation}
so the limit values are exact : at criticality $\lambda_{q}(0)$ is by
definition equal to the critical exponent : $\gamma$, $\nu$ or $D\nu$
for the susceptibility, the reduced correlation length, and the
normalized Binder cumulant respectively. Close to criticality the
leading confluent correction amplitude can be estimated from the
initial slope of the $\lambda_{q}(\tau)$ against $\tau^{\theta}$ plot,
as \cite{butera:02}
\begin{equation}
  \lambda_{q}(\tau) = -\partial\ln Q(\tau)/\partial\ln\tau =
  \lambda_{q} - a_{q}\theta\tau^{\theta}
\end{equation}
In practice this limiting slope is hard to estimate accurately. As a
result the values of the universal ratios such as $a_{\chi}/a_{\xi}$
are only known approximately.  For the normalized Binder cumulant
$a_{q}=0$ (see below) in which case
\begin{equation}
  \lambda_{q}(\tau) = -\partial\ln Q(\tau)/\partial\ln\tau =
  \lambda_{q} - b_{q}\theta\tau
\end{equation} 

With the three correction-term expression above, one has
at infinite temperature the limiting value
\begin{equation}
  \lambda_{q}(\tau=1) = \lambda_{q} - \frac{a_{q}\theta + b_{q} +
    c_{q}\mu}{1 + a_{q} + b_{q} + c_{q}}
\end{equation}
At infinite temperature, from the leading HTSE series
terms~\cite{butera:02a} one also knows that the $S=1/2$,
$\lambda_{q}(\tau=1)$ limiting values are equal to $z\beta_{c}$,
$(z/2)\beta_{c}$ and $2 z \beta_{c}$ for the susceptibility, the
reduced correlation length, and the normalized Binder cumulant
respectively where $z$ is the number of near neighbors. These exact
limit values are indicated by red arrows in each of the effective
exponent plots. These two relations provide an additional closure
condition for each observable on the fit correction-term amplitudes
together with the fit value for the exponent $\mu$. In particular for
the normalized Binder cumulant data with $a_{g}=0$ (see below) the
parameters $C_{g}$ and $b_{g}$ can be read off the critical limit
plots; then as $C_{g}(1+b_{g}+c{g})= 1$, the infinite temperature
limit condition
\begin{equation}
  2 z \beta_{c} = 3 \nu - \frac{b_{g}+c_{g}\mu}{1 + b_{g} + c_{g}}
\end{equation}
leaves $\mu$ fixed.   

\section{Face Centered Cubic lattice}\label{sec:IV}
\begin{figure}
  \includegraphics[width=3.4in]{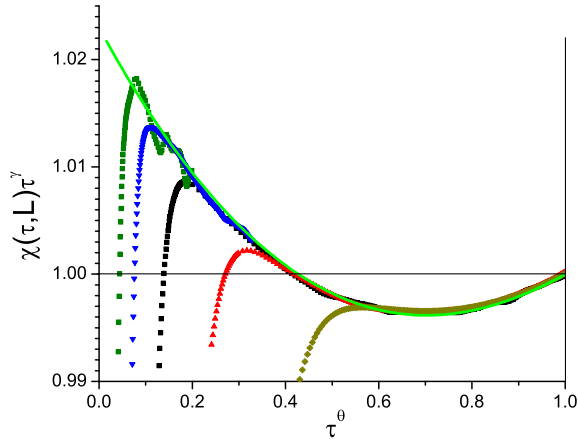}
  \caption{(Color on line) FCC lattice. Susceptibility data in the
    form $\chi(\tau,L)\tau^{\gamma}$ against $\tau^{\theta}$. Data for
    $L= 128$, $64$, $32$, $16$, $8$ from left to right. Green curve is
    the fit to the ThL data, see Eq.~\eqref{fccchifit}.}
  \protect\label{fig1}
\end{figure}

\begin{figure}
  \includegraphics[width=3.4in]{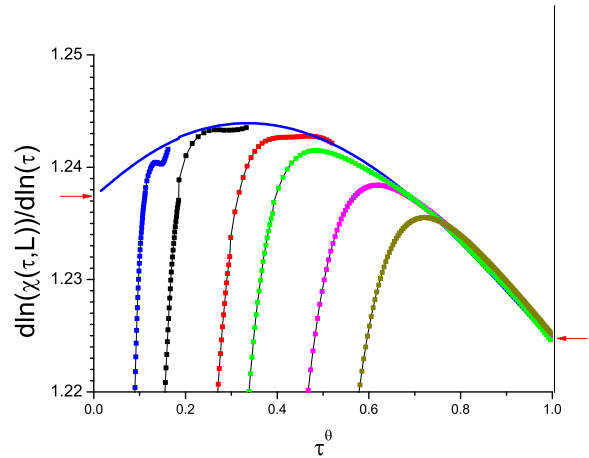}
  \caption{(Color on line) FCC lattice. The temperature dependent
    effective susceptibility exponent
    $\partial\ln\chi(\tau,L)/\partial\ln\tau$ against
    $\tau^{\theta}$. Data for $L= 64$, $32$, $16$, $12$, $8$, $6$ from
    left to right. Green curve is the fit to the ThL data, calculated
    from $\chi(\tau)$ in Eq.~\eqref{fccchifit}.}  \protect\label{fig2}
\end{figure}

In this lattice each site has 12 near neighbors and 4 sites per unit
cell. The critical inverse temperature is $\beta_c = 0.102069(1)$
\cite{lundow:09,murase:08}. The ThL critical amplitudes for the
susceptibility and the normalized correlation length $\chi(\tau,L)$
and $\xi(\tau)/\beta^{1/2}$ are both close to $1$. The susceptibility
data can be fitted satisfactorily with three weak correction terms
only : the leading confluent correction $a\tau^{\theta}$, the leading
analytic correction $b\tau$ and the further term $c\tau^{\mu}$
\begin{equation}
  \chi(\tau) = 1.023\tau^{-\gamma}\left(1 -0.080\tau^{\theta} +0.0595\tau
  -0.0022\tau^{\mu}\right)
  \label{fccchifit}
\end{equation}
with $\mu \sim 2.45$, see Figs.~\ref{fig1} and \ref{fig2}.

\begin{figure}
  \includegraphics[width=3.4in]{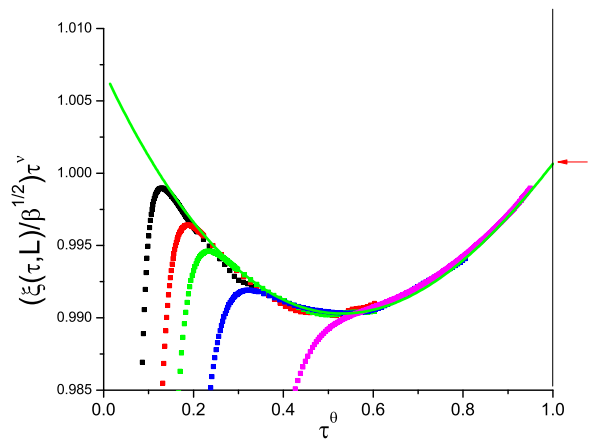}
  \caption{(Color on line) FCC lattice. Regularized correlation length
    data in the form $[\xi(\tau,L)/\beta^{1/2}]\tau^{\nu}$ against
    $\tau^{\theta}$. Data for $L= 48$, $32$, $24$, $16$, $8$ from left
    to right. Green curve is the fit to the ThL data, see
    Eq.~\eqref{fccxifit}.}  \protect\label{fig3}
\end{figure}

\begin{figure}
  \includegraphics[width=3.4in]{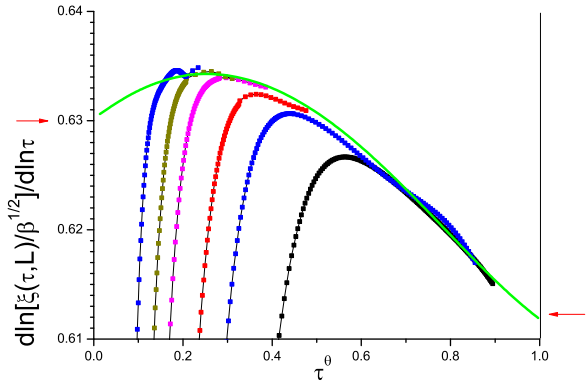}
  \caption{(Color on line) FCC lattice. The temperature dependent
    effective correlation length exponent
    $\partial\ln[\xi(\tau,L)/\beta^{1/2}]/\partial\ln\tau$ against
    $\tau^{\theta}$. Data for $L= 48$, $32$, $24$, $16$, $12$, $8$
    from left to right. Green curve is the fit to the ThL data,
    calculated from Eq.~\eqref{fccxifit}.}  \protect\label{fig4}
\end{figure}

The reduced correlation length in the ThL regime can also be fitted
with three terms only :
\begin{equation}
  \frac{\xi(\tau)}{\beta^{1/2}} = 1.0071
  \tau^{-\nu}\left(1-0.0655\tau^{\theta}+0.0635\tau
  -0.0043\tau^{\mu}\right)
  \label{fccxifit}
\end{equation}
again with $\mu \sim 2.45$, see Figs.~\ref{fig3} and \ref{fig4}.  The
effective exponents $\gamma(\tau)$ and $\nu(\tau)$ each vary by only
about $1\%$ over the entire temperature range from criticality to
infinity.  The value $\mu \sim 2.45$ for the tiny third correction term
exponents is only rough.

The results for the normalized Binder parameter are more
remarkable. The usual leading confluent correction turns out to have
zero amplitude and the only visible correction term is the analytic
term which is strong and linear in $\tau$. All further higher-order
correction terms have negligible amplitudes also, so
\begin{equation}
  L^{3}g(\tau,L) = 1.614 \tau^{-3\nu}\left(1 - 0.380\tau\right)
  \label{fccL3gfit}
\end{equation}

As the normalized Binder cumulant is equal to
$-\chi_{4}(\tau)/2\chi(\tau)^2$, the absence of the leading confluent
correction term implies that the $\chi_{4}(\tau)$ and $\chi(\tau)$
confluent correction amplitudes have a ratio $a_{\chi_4}/a_{\chi} =
2$. Because confluent correction amplitude ratios $a(Q_{i})/a(Q_{j})$
are universal \cite{ferer:77,chang:80}, the normalized Binder
parameter leading confluent correction amplitude will be zero for all
models in the 3D Ising universality class. This is indeed confirmed
below from the data for the other models studied.

\begin{figure}
  \includegraphics[width=3.4in]{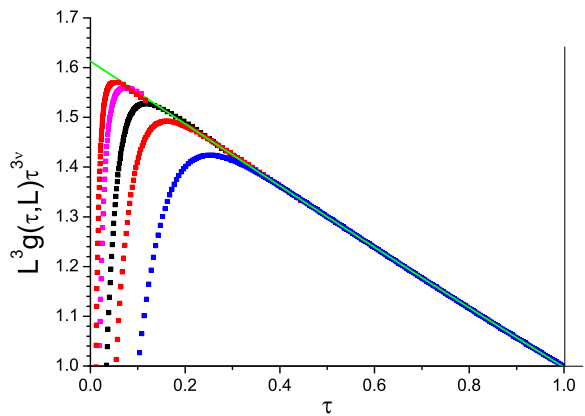}
  \caption{(Color on line) FCC lattice. Normalized Binder cumulant
    data in the form $L^{3}g(\tau,L)\tau^{3\nu}$ against $\tau$. Data
    for $L= 32$, $24$, $16$, $12$, $8$ from left to right. Green curve
    is the fit to the ThL data, see Eq.~\eqref{fccL3gfit}.}
  \protect\label{fig5}
\end{figure}

\begin{figure}
  \includegraphics[width=3.4in]{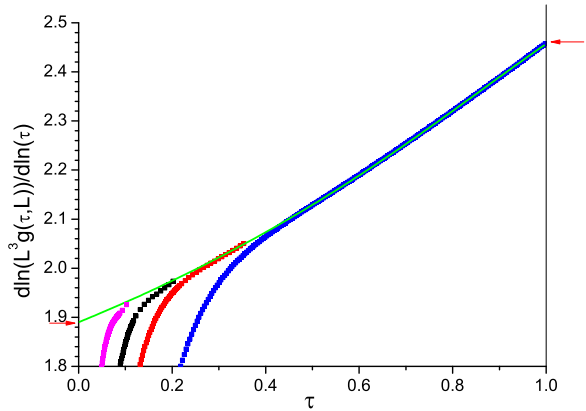}
  \caption{(Color on line) FCC lattice. The temperature dependent
    effective normalized Binder cumulant exponent in the form
    $\partial\ln[L^{3}g(\tau,L)]/\partial\ln\tau$ against $\tau$. Data
    for $L= 24$, $16$, $12$, $8$ from left to right. Green curve is
    the fit to the ThL data, calculated from Eq.~\eqref{fccL3gfit}.}
  \protect\label{fig6}
\end{figure}

\section{Body Centered Cubic lattice}\label{sec:V}
\begin{figure}
  \includegraphics[width=3.4in]{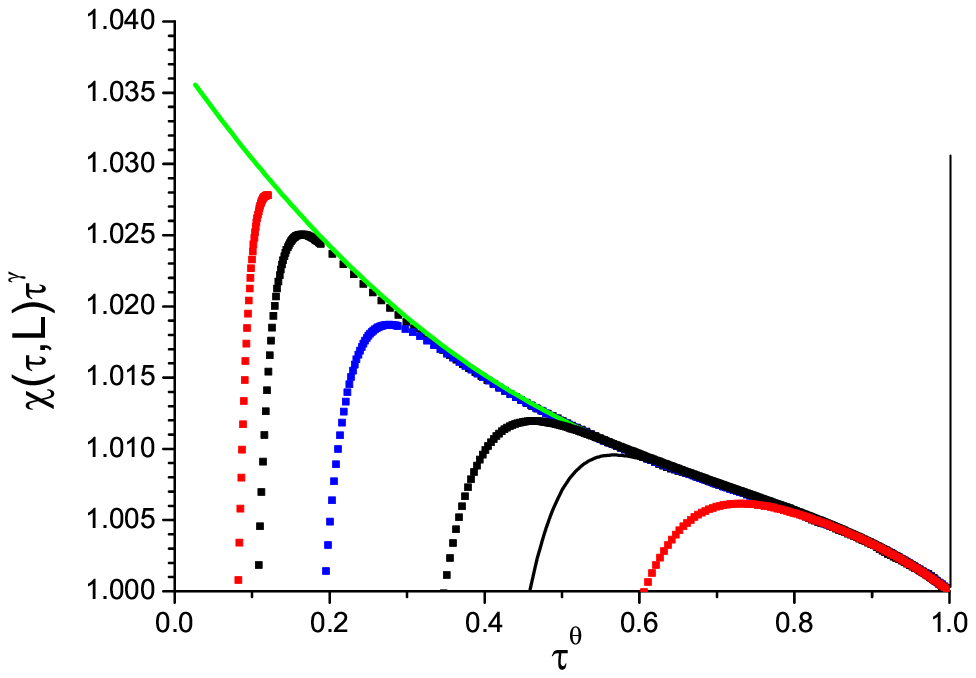}
  \caption{(Color on line) BCC lattice. Susceptibility data in the
    form $\chi(\tau,L)\tau^{\gamma}$ against $\tau^{\theta}$. Data for
    $L= 96$, $64$, $48$, $24$, $12$, HTSE and $6$ from left to
    right. The HTSE curve is a 23 term sum of data from
    Ref.~\cite{butera:02a}. Green curve is the fit to the ThL data,
    see Eq.~\eqref{bccchifit}.}  \protect\label{fig7}
\end{figure}

\begin{figure}
  \includegraphics[width=3.4in]{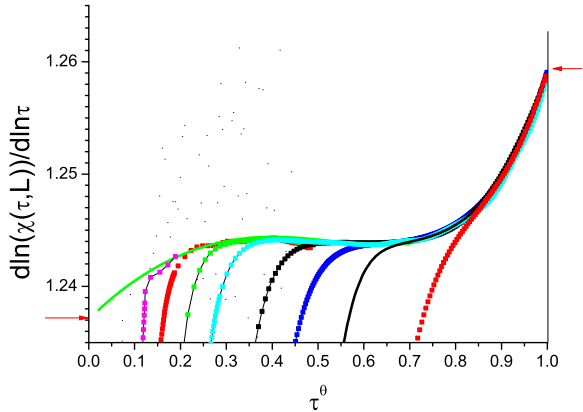}
  \caption{(Color on line) BCC lattice. The temperature dependent
    effective susceptibility exponent
    $\partial\ln\chi(\tau,L)/\partial\ln\tau$ against
    $\tau^{\theta}$. Data for $L= 64$, $48$, $32$, $24$, $16$, $12$,
    HTSE and $6$ from left to right. The HTSE curve is a 23 term sum
    of data from Ref.~\cite{butera:02a}.  Green curve is the fit to
    the ThL data, calculated from $\chi(\tau)$ in
    Eq.~\eqref{bccchifit}.}  \protect\label{fig8}
\end{figure}

In this lattice each site has 8 near neighbors and 2 sites per unit
cell. The critical inverse temperature $\beta_c = 0.1573725(5)$
\cite{lundow:09,murase:08,butera:02}. Extensive lists of exact HTSE
terms for this lattice are given in Ref.~\cite{butera:02a}; we have
used these tables to calculate HTSE values for the observables in the
high-temperature range where the HTSE sums are essentially exact.
Accurate effective exponents to lower temperatures can be obtained by
appropriate extrapolation (see Ref.~\cite{butera:02}).  The ThL
susceptibility $\chi(\tau,L)$ from simulations and HTSE data can be
fitted satisfactorily with three correction terms :
\begin{equation}
  \chi(\tau) = 1.0377\tau^{-\gamma}\left(1 -0.0771\tau^{\theta} +0.054\tau
    -0.0137\tau^{\mu}\right)
  \label{bccchifit}
\end{equation}
with $\mu \sim 2.45$, see Figs.~\ref{fig7} and \ref{fig8}. The latter
is essentially identical to the curve shown in Ref.~\cite{butera:02},
Fig.~14.

\begin{figure}
  \includegraphics[width=3.4in]{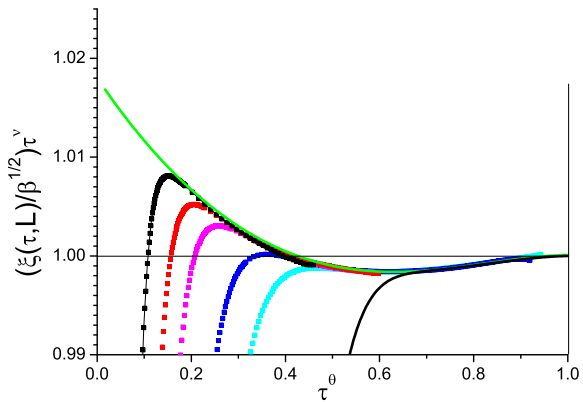}
  \caption{(Color on line) BCC lattice. Reduced correlation length
    data in the form $[\xi(\tau,L)/\beta^{1/2}]\tau^{\nu}$ against
    $\tau^{\theta}$. Data for $L= 48$, $32$ $24$, $12$ and HTSE from
    left to right. The HTSE curve is a 23 term sum of data from
    Ref.~\cite{butera:02a}. Green curve is the fit to the ThL data,
    see Eq.~\eqref{bccxifit}.}  \protect\label{fig9}
\end{figure}

\begin{figure}
  \includegraphics[width=3.4in]{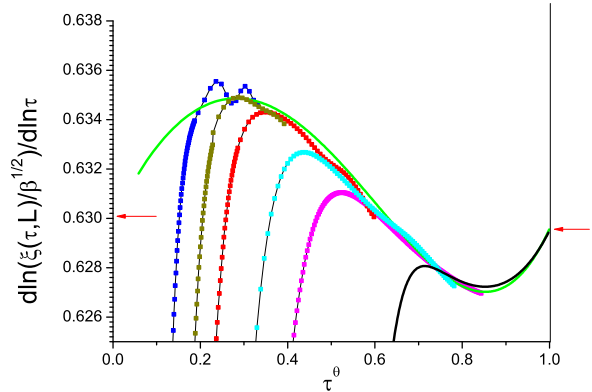}
  \caption{(Color on line) BCC lattice. The temperature dependent
    effective correlation length exponent
    $\partial\ln[\xi(\tau,L)/\beta^{1/2}]/\partial\ln\tau$ against
    $\tau^{\theta}$. Data for $L= 48$, $32$, $24$, $16$, $12$ and HTSE
    from left to right. The HTSE curve is a 23 term sum of data from
    Ref.~\cite{butera:02a}. Green curve is the fit to the ThL data,
    calculated from Eq.~\eqref{bccxifit}.}  \protect\label{fig10}
\end{figure}

The fit values for the critical amplitude and the confluent correction
amplitude can be compared to those estimated in Ref.~\cite{butera:02},
$C_{\chi} = 1.0404(1), a_{\chi} = -0.129(3)$.  Because of the opposite
signs of the various correction term amplitudes the temperature
dependent effective exponent $\gamma(\tau)$ changes slope twice. The
upturns in $\gamma(\tau)$ at high temperatures in both plots
correspond to the weak but not negligible negative amplitude
correction term with exponent $\mu \sim 2.45$.  The reduced
correlation length in the ThL regime can also be fitted with three
terms :
\begin{equation}
  \frac{\xi(\tau)}{\beta^{1/2}} =
  1.018\tau^{-\nu}\left(1-0.069\tau^{\theta}+ 0.0619 \tau
  -0.0102\tau^{\mu}\right)
  \label{bccxifit}
\end{equation}
with $\mu \sim 2.45$, see Figs.~\ref{fig9} and \ref{fig10}.

\begin{figure}
  \includegraphics[width=3.4in]{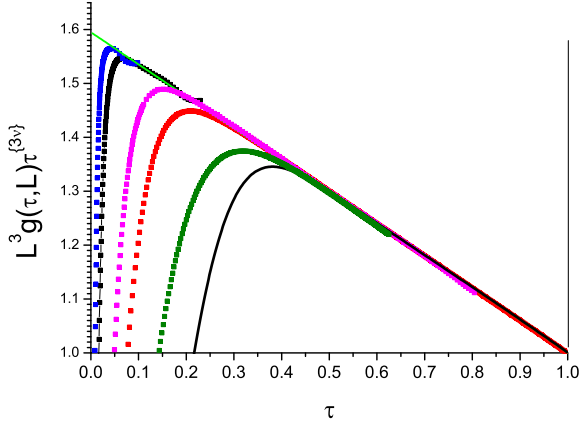}
  \caption{(Color on line) BCC lattice. Normalized Binder cumulant
    data in the form $L^{3}g(\tau,L)\tau^{3\nu}$ against $\tau$. Data
    for $L= 48$, $32$, $16$, $12$, $8$ and HTSE from left to
    right. The HTSE curve is a 23 term sum of data from
    Ref.~\cite{butera:02a}. Green curve is the fit to the ThL data,
    see Eq.~\eqref{fccL3gfit}.}  \protect\label{fig11}
\end{figure}

\begin{figure}
  \includegraphics[width=3.4in]{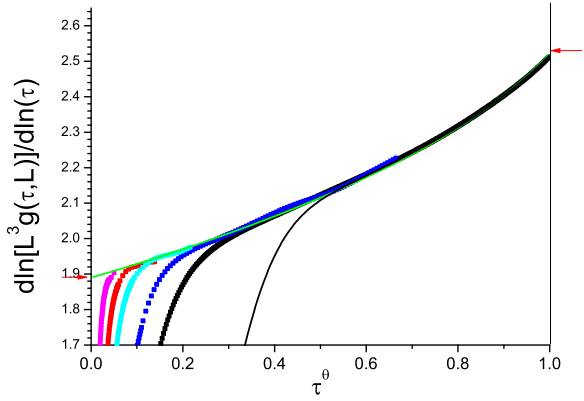}
  \caption{(Color on line) BCC lattice. The temperature dependent
    effective normalized Binder cumulant exponent in the form
    $\partial\ln[L^{3}g(\tau,L)]/\partial\ln\tau$ against $\tau$. Data
    for $L= 48$, $32$, $24$, $16$, $12$ and HTSE from left to
    right. The HTSE curve is a 23 term sum of data from
    Ref.~\cite{butera:02a}. Green curve is the fit to the ThL data,
    calculated from Eq.~\eqref{bccL3gfit}.}  \protect\label{fig12}
\end{figure}

The present reduced correlation length critical amplitude $C_{\xi}$
corresponds to a conventional critical amplitude
$C_{\xi}\beta_c^{1/2}= 0.404$.  The critical amplitude and confluent
correction amplitude can be compared to estimates $0.4681(3)$ and
$-0.100(4)$ in Ref.~\cite{butera:02}.  The ThL normalized Binder
parameter behaves slightly differently from the FCC model. The
dominant correction term is again the strong analytic term linear in
$\tau$; however there is a further weak term having an exponent $\mu
\sim 2.45$. All other terms are missing including the normal leading
correction term in $\tau^{\theta}$ and "minor" correction terms, so :
\begin{equation}
  L^{3}g(\tau,L) = 1.597 \tau^{-3\nu}\left(1 - 0.3657\tau -
  0.0068\tau^{\mu}\right)
  \label{bccL3gfit}
\end{equation}
with $\mu \sim 2.65$, see Figs.~\ref{fig11} and \ref{fig12}.

\section{Simple Cubic lattice}\label{sec:VI}
\begin{figure}
  \includegraphics[width=3.4in]{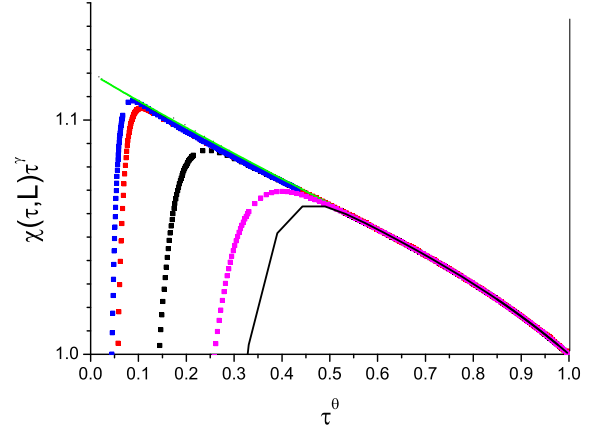}
  \caption{(Color on line) SC lattice. Susceptibility data in the form
    $\chi(\tau,L)\tau^{\gamma}$ against $\tau^{\theta}$. Data for
    $L=64$, $48$, $16$, $8$ and HTSE from left to right. The HTSE
    curve is a 23 term sum of data from Ref.~\cite{butera:02a}. Green
    curve is the fit to the ThL data, see Eq.~\eqref{scchifit}.}
  \protect\label{fig13}
\end{figure}

\begin{figure}
  \includegraphics[width=3.4in]{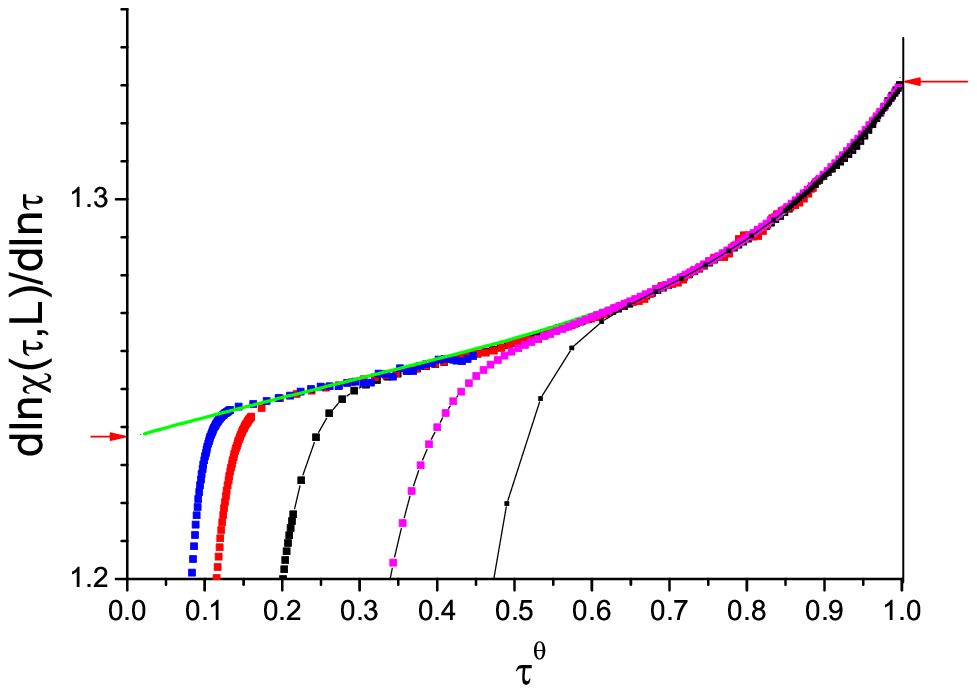}
  \caption{(Color on line) SC lattice. The temperature dependent
    effective susceptibility exponent
    $\partial\ln\chi(\tau,L)/\partial\ln\tau$ against
    $\tau^{\theta}$. Data for $L= 48$, $32$, $16$, $8$ and HTSE from
    left to right. The HTSE curve is a 23 term sum of data from
    Ref.~\cite{butera:02a}.  Green curve is the fit to the ThL data,
    calculated from $\chi(\tau)$ in Eq.~\eqref{scchifit}.}
  \protect\label{fig14}
\end{figure}

In this lattice each site has 6 near neighbors and 1 site per unit
cell. The critical inverse temperature is $\beta_c = 0.221654(2)$
\cite{blote:95,haggkvist:07,butera:02}. Extensive lists of exact HTSE
terms for this lattice are given in Ref.~\cite{butera:02a}. The ThL
susceptibility $\chi(\tau)$ data and the high-temperature HTSE data
can be fitted satisfactorily with three correction terms :
\begin{equation}
  \chi(\tau) = 1.120\tau^{-\gamma}\left(1 -0.112\tau^{\theta}
  +0.021\tau-0.019\tau^{\mu}\right)
  \label{scchifit}
\end{equation}
with $\mu \sim 2.45$, see Figs.~\ref{fig13} and \ref{fig14}. The fit
value for the critical amplitude can be compared to that estimated in
Ref.~\cite{butera:02}, $C_{\chi} = 1.14(1)$. The temperature-dependent
ThL effective exponent $\gamma(\tau)$ in Fig.~\ref{fig14} is very
similar to the $\gamma(\tau)$ curve for the same model shown in
Ref.~\cite{butera:02} Fig.~16.

\begin{figure}
  \includegraphics[width=3.4in]{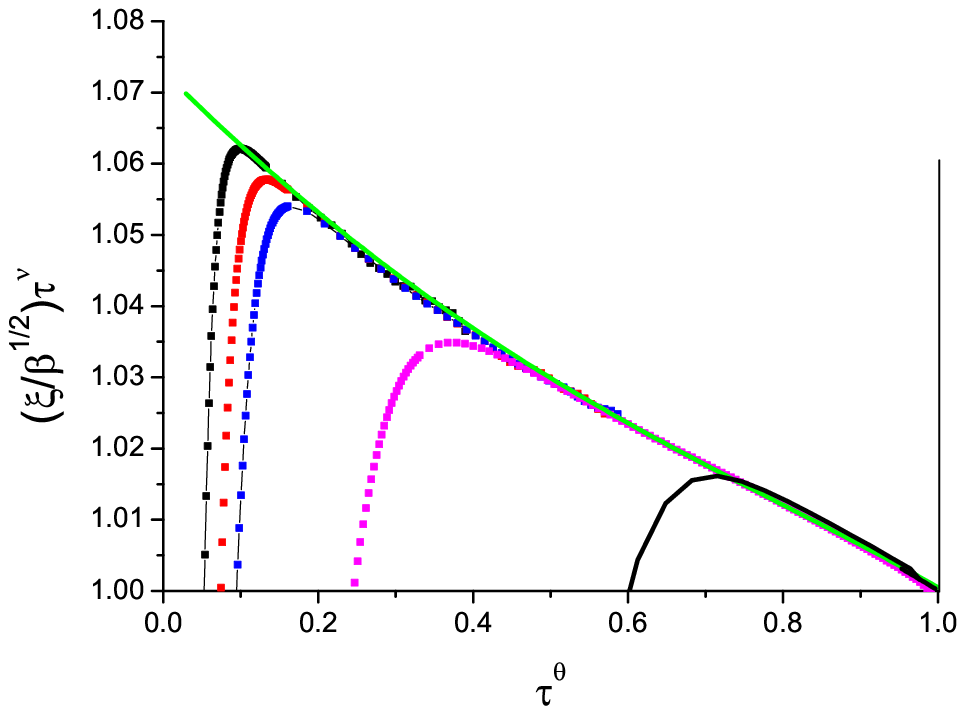}
  \caption{(Color on line) SC lattice. Normalized correlation length
    data in the form $[\xi(\tau,L)/\beta^{1/2}]\tau^{\nu}$ against
    $\tau^{\theta}$. Data for $L= 48$, $32$, $16$, $8$, HTSE from left
    to right. The HTSE curve is a 23 term sum of data from
    Ref.~\cite{butera:02a}.  Green curve is the fit to the ThL data,
    see Eq.~\eqref{scxifit}.}  \protect\label{fig15}
\end{figure}

\begin{figure}
  \includegraphics[width=3.4in]{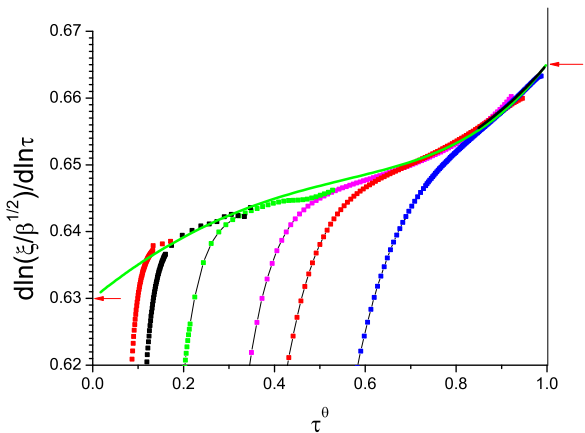}
  \caption{(Color on line) SC lattice. The temperature dependent
    effective correlation length exponent
    $\partial\ln[\xi(\tau,L)/\beta^{1/2}]/\partial\ln\tau$ against
    $\tau^{\theta}$. Data for $L= 48$, $32$, $16$, $12$, $8$, $6$, $4$
    from left to right. Green curve is the fit to the ThL data,
    calculated from Eq.~\eqref{fccxifit}.}  \protect\label{fig16}
\end{figure}

The reduced correlation length in the ThL regime can be fitted by
\begin{equation}
  \frac{\xi(\tau)}{\beta^{1/2}} = 1.073
  \tau^{-\nu}\left(1-0.107\tau^{\theta}+0.048\tau -
  0.010\tau^{\mu}\right)
  \label{scxifit}
\end{equation}
with $\mu \sim 2.45$, see Figs.~\ref{fig15} and \ref{fig16}.  The fit
value for the critical amplitude can be compared to that estimated in
Ref.~\cite{butera:02}, equivalent to $C_{\xi} = 1.077(12)$.

The ratios $a_{\chi}/a_{\xi}$ should be identical for these three
models. The values estimated above are $1.22$ for the FCC model,
$1.32$ for the BCC and $1.05$ for the SC. The BCC model values for
different spins $S$ estimated in Ref.~\cite{butera:02} were all close
to $1.28$. The present variations reflect the difficulties in
extrapolating precisely so as to estimate the initial critical slopes.

\begin{figure}
  \includegraphics[width=3.4in]{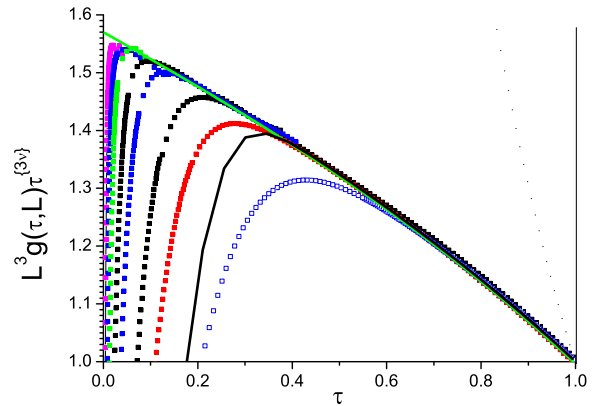}
  \caption{(Color on line) SC lattice. Normalized Binder cumulant data
    in the form $L^{3}g(\tau,L)\tau^{3\nu}$ against $\tau$. Data for
    $L= 48$, $32$, $16$, $12$, $8$, $6$, HTSE and $4$ from left to
    right. The HTSE curve is a 23 term sum of data from
    Ref.~\cite{butera:02a}. Green curve is the fit to the ThL data,
    see Eq.~\eqref{scL3gfit}.}  \protect\label{fig17}
\end{figure}

\begin{figure}
  \includegraphics[width=3.4in]{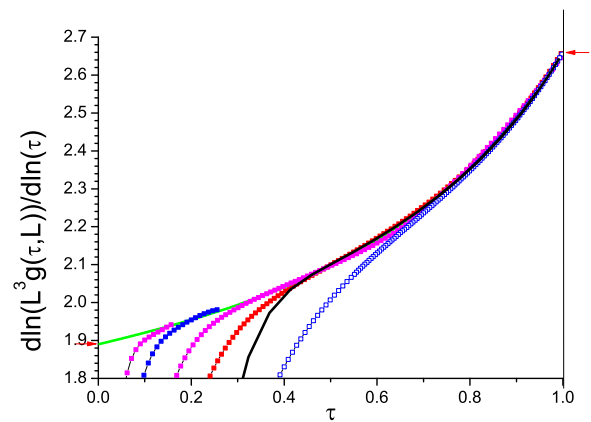}
  \caption{(Color on line) SC lattice. The temperature dependent
    effective normalized Binder cumulant exponent in the form
    $\partial\ln[L^{3}g(\tau,L)]/\partial\ln\tau$ against $\tau$. Data
    for $L= 16$, $12$, $8$, $6$, HTSE and $4$ from left to right. The
    HTSE curve is a 23 term sum of data from
    Ref.~\cite{butera:02a}. Green curve is the fit to the ThL data,
    calculated from Eq.~\eqref{scL3gfit}.}  \protect\label{fig18}
\end{figure}

For the normalized Binder parameter, as for the other lattices the
standard leading correction term in $\tau^{\theta}$ is missing. There
is a strong analytic correction term linear in $\tau$, accompanied by
another strong term proportional to $\tau^{\mu}$ with $\mu$ close to
$2.45$. All other terms are negligible so that
\begin{equation}
  L^{3}g(\tau,L) = 1.565 \tau^{-3\nu}\left(1 - 0.282\tau - 0.081
  \tau^{\mu}\right)
  \label{scL3gfit}
\end{equation}
with $\mu \sim 2.65$, see Figs.~\ref{fig17} and \ref{fig18}.

We identify this $\mu$ with the subleading confluent correction
exponent, so we estimate $\theta_{2} = 2.45(5)$. All the fits to the
data sets for this and the other models are compatible with a
universal correction term being present having approximately this
exponent. Obviously when the high exponent correction term amplitude
is very weak, as the case for instance for the SC $\chi(\tau)$ and
$\xi(\tau)/\beta^{1/2}$ data sets, the estimate for the corresponding
$\mu$ is much more approximate. Nevertheless acceptable fits to these
data sets also can only be made when a high-exponent correction term
is included.

\section{Diamond lattice}\label{sec:VII}
\begin{figure}
  \includegraphics[width=3.4in]{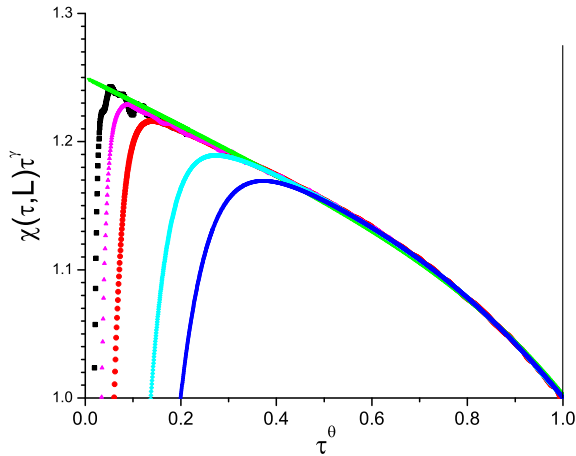}
  \caption{(Color on line) Diamond lattice. Susceptibility data in the
    form $\chi(\tau,L)\tau^{\gamma}$ against $\tau^{\theta}$. Data for
    $L= 128$, $64$, $32$, $12$, $8$ from left to right. Green curve is
    the fit to the ThL data, see Eq.~\eqref{diamchifit}.}
  \protect\label{fig19}
\end{figure}

\begin{figure}
  \includegraphics[width=3.4in]{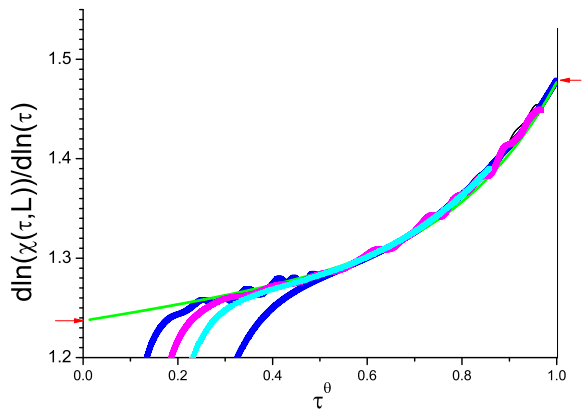}
  \caption{(Color on line) Diamond lattice. The temperature dependent
    effective susceptibility exponent
    $\partial\ln\chi(\tau,L)/\partial\ln\tau$ against
    $\tau^{\theta}$. Data for $L= 24$, $16$, $12$, $8$ from left to
    right. Green curve is the fit to the ThL data, calculated from
    $\chi(\tau)$ in Eq.~\eqref{diamchifit}.}  \protect\label{fig20}
\end{figure}

In this lattice each site has 4 near neighbors and 8 sites per unit
cell. The critical inverse temperature $\beta_c = 0.3697398(1)$
\cite{lundow:09,deng:03}. For technical reasons it is more difficult
to equilibrate and obtain accurate numerical data for this model.  The
ThL susceptibility $\chi(\tau)$ can be fitted satisfactorily with
three correction terms :
\begin{equation}
  \chi(\tau) = 1.250\tau^{-\gamma}\left(1-0.147\tau^{\theta} -
  0.011\tau - 0.04\tau^{\mu}\right)
  \label{diamchifit}
\end{equation}
with $\mu \sim 2.45$, see Figs.~\ref{fig19} and \ref{fig20}.  We do not
dispose of sufficient data to analyse the normalized correlation
length in this model.

\begin{figure}
  \includegraphics[width=3.4in]{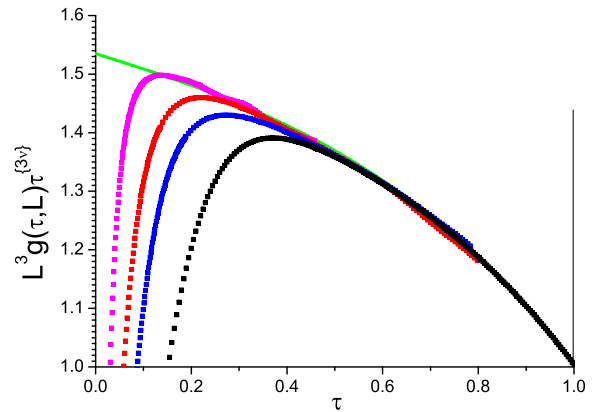}
  \caption{(Color on line) Diamond lattice. Normalized Binder cumulant
    data in the form $L^{3}g(\tau,L)\tau^{3\nu}$ against $\tau$. Data
    for $L= 12$, $8$, $6$, $4$ from left to right. Green curve is the
    fit to the ThL data, see Eq.~\eqref{diamL3gfit}.}
  \protect\label{fig21}
\end{figure}

\begin{figure}
  \includegraphics[width=3.4in]{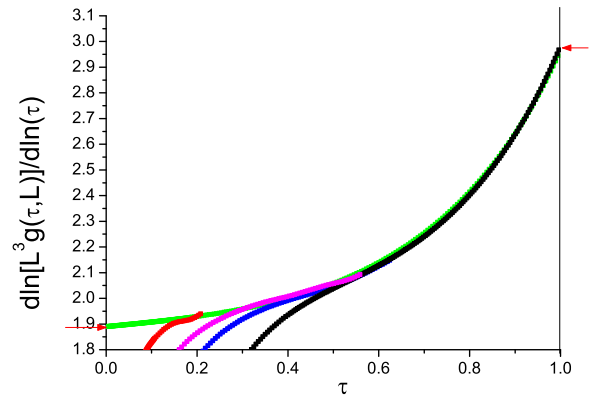}
  \caption{(Color on line) Diamond lattice. The temperature dependent
    effective normalized Binder cumulant exponent in the form
    $\partial\ln[L^{3}g(\tau,L)]/\partial\ln\tau$ against $\tau$. Data
    for $L= 12$, $8$, $6$, $4$ from left to right. Green curve is the
    fit to the ThL data, calculated from Eq.~\eqref{diamL3gfit}.}
  \protect\label{fig22}
\end{figure}

The normalized Binder parameter behaves in much the same way as in the
SC model. The leading correction term is strong and linear in $\tau$,
with a strong second term proportional to $\tau^{\mu}$ with $\mu \sim
2.7$ so
\begin{equation}
  L^{3}g(\tau,L) = 1.535 \tau^{-3\nu}\left(1 - 0.157\tau - 0.186
  \tau^{\mu}\right),
  \label{diamL3gfit}
\end{equation}
see Figs.~\ref{fig21} and \ref{fig22}.  The $\mu$ correction term
amplitude is even stronger than for the SC model. Unfortunately no
HTSE data are available for this model. The estimate for $\mu$ is
marginally higher than the SC and BCC model Binder cumulant analyses
but this may be due to technical difficulties with this model.

\section{Conclusion}\label{sec:VIII}
We measure the susceptibility, reduced second-moment correlation
length, and normalized Binder-cumulant data for the 3D spin-$1/2$ FCC,
BCC, SC and diamond Ising models, covering the entire paramagnetic
temperature range. We treat the bootstrap values for the principle
critical exponents as exact and carry out three term (or two term for
the normalized Binder cumulant) fits adjusting the critical amplitudes
and the correction-term amplitudes, including a high-order term with
exponent approximately equal to the bootstrap $\theta_{2}$ value. Our
principal conclusion is that for the models and observables studied,
there systematically exist correction terms of exponent consistent
with the bootstrap subleading conformal correction term value
$\theta_{2}=2.454(3)$ \cite{simmons:17}, and with significant
amplitudes. For all three observables these high order correction term
$c\tau^{2.45}$ amplitudes are always negative and pass progressively
from almost negligible for the FCC lattice to strong for the SC and
Diamond lattices. This evolution is particularly notable for the
normalized Binder cumulant.

All the critical amplitudes and correction-term amplitudes evolve
regularly from one model to the next as functions of the numbers of
nearest neighbors, with the susceptibility and correlation-length
critical amplitudes becoming systematically stronger as the number of
neighbors drops.  Amplitude ratios for the leading confluent
correction terms for different observables such as $a_{\chi}/a_{\xi}$
are universal. Our estimates for this ratio are broadly compatible
with a value $\sim 1.25$ \cite{butera:02} (though the value of this
particular ratio turns out to be hard to estimate accurately).

The amplitude ratio universality implies that if for an observable
$Q(\tau)$, $a_{q} = 0$ for one particular model then $a_{q}$ must also
be equal zero for all other models in the same university class. The
data show that this rule is indeed obeyed for the normalized Binder
parameter in all four models studied; the leading confluent correction
term $a_{g}\tau^{\theta}$ is absent to within the statistical
uncertainty. As the normalized Binder parameter is equal to
$-\chi_{4}(\tau)/(2\chi(\tau)^2)$ the correction Binder-cumulant
amplitude ratio being equal to zero is equivalent to
$a_{\chi_{4}}/a_{\chi} = 2$ for the 3D Ising universality class.  The
normalized Binder cumulant analytic corrections $b\tau$ are always
strong but decrease progressively as the number of neighbors drops.
All "minor" correction term amplitudes appear to be negligible in all
cases.

\begin{acknowledgments}
  We would like to thank D.~Simmons-Duffin, P.~Butera, S.~Rychkov and
  Y.~Nakayama for helpful comments.  The computations were performed
  on resources provided by the Swedish National Infrastructure for
  Computing (SNIC) at Chalmers Centre for Computational Science and
  Engineering (C3SE).
\end{acknowledgments}

\end{document}